\documentclass[letter,bibyear]{aa} 

\usepackage{epsfig}
\usepackage{amsmath}
\usepackage{subfigure}
\usepackage{natbib}
\usepackage{multirow}
\usepackage{color}
\usepackage{longtable}
\usepackage{graphicx}
\usepackage{epstopdf}
\usepackage{booktabs}
\usepackage{txfonts}

\newcommand{\jms}{J.~Mol.~Spectrosc.}

\newcommand{\once}{10$^{11}$\,cm$^{-2}$}

\begin{document}

\title{Discovery of  CH$_3$CHCO in TMC-1 with the QUIJOTE line survey\thanks{Based on 
observations carried out
with the Yebes 40m telescope (projects 19A003,
20A014, 20D023, 21A011, and 21D005). The 40m
radiotelescope at Yebes Observatory is operated by the Spanish Geographic Institute
(IGN, Ministerio de Transportes, Movilidad y Agenda Urbana).}}

\author{
R.~Fuentetaja\inst{1},
C~.Berm\'udez\inst{2},
C.~Cabezas\inst{1},
M.~Ag\'undez\inst{1},
B.~Tercero\inst{3,4},
N.~Marcelino\inst{3,4},
J.~R.~Pardo\inst{1},
L.~Margul\`es\inst{5},
R.~A.~Motiyenko\inst{5},
J.~-C.~Guillemin\inst{6},
P.~de~Vicente\inst{3},
J.~Cernicharo\inst{1}
}

\institute{Dept. de Astrof\'isica Molecular, Instituto de F\'isica Fundamental (IFF-CSIC),
C/ Serrano 121, 28006 Madrid, Spain. \email r.fuentetaja@csic.es, jose.cernicharo@csic.es
\and Dept. de Química Física y Química Inorgánica, Facultad de Ciencias - I.U. CINQUIMA, Universidad de Valladolid, Paseo de Belén 7, 47011 Valladolid, Spain.
\and Centro de Desarrollos Tecnol\'ogicos, Observatorio de Yebes (IGN), 19141 Yebes, Guadalajara, Spain.
\and Observatorio Astron\'omico Nacional (OAN, IGN), C/ Alfonso XII, 3, 28014, Madrid, Spain.
\and Universit\'e de Lille, Facult\'e des Sciences et Technologies, D\'epartement Physique, Laboratoire de Physique des Lasers, Atomes et Molécules, UMR CNRS 8523, 59655 Villeneuve d’Ascq Cedex, France.
\and Univ. Rennes, Ecole Nationale Sup\'erieure de Chimie de Rennes, CNRS, ISCR – UMR6226, 35000 Rennes, France.
}

\date{}

\abstract{We report the detection of methyl ketene towards TMC-1 with the QUIJOTE line survey. Nineteen rotational transitions with rotational quantum numbers ranging from $J$ = 3 up to $J$ = 5 and $K_a$ $\le$ 2 were identified in the frequency range 32.0-50.4 GHz, 11 of which arise above the 3$\sigma$ level. 
We derived a column density for CH$_3$CHCO of $N$=1.5$\times$10$^{11}$ cm$^{-2}$ and a rotational temperature of 9 K. Hence, the abundance ratio between ketene and methyl ketene, CH$_2$CO/CH$_3$CHCO, is 93. This species is the second C$_3$H$_4$O isomer detected. The other, $trans$-propenal (CH$_2$CHCHO), corresponds to the most stable isomer and has a column density of $N$=(2.2$\pm$0.3)$\times$10$^{11}$ cm$^{-2}$, which results in an abundance ratio CH$_2$CHCHO/CH$_3$CHCO of 1.5. The next non-detected isomer with the lowest energy is   $cis$-propenal, which  is  therefore a good candidate for future discovery. We have carried out an in-depth study of the possible gas-phase chemical reactions involving methyl ketene to explain the abundance detected, achieving good agreement between chemical models and observations.}


\titlerunning{CH3CHCO in TMC-1}
\authorrunning{Fuentetaja et al.}

\maketitle

\begin{figure*}[htp]
\centering
\includegraphics[width=0.8\textwidth]{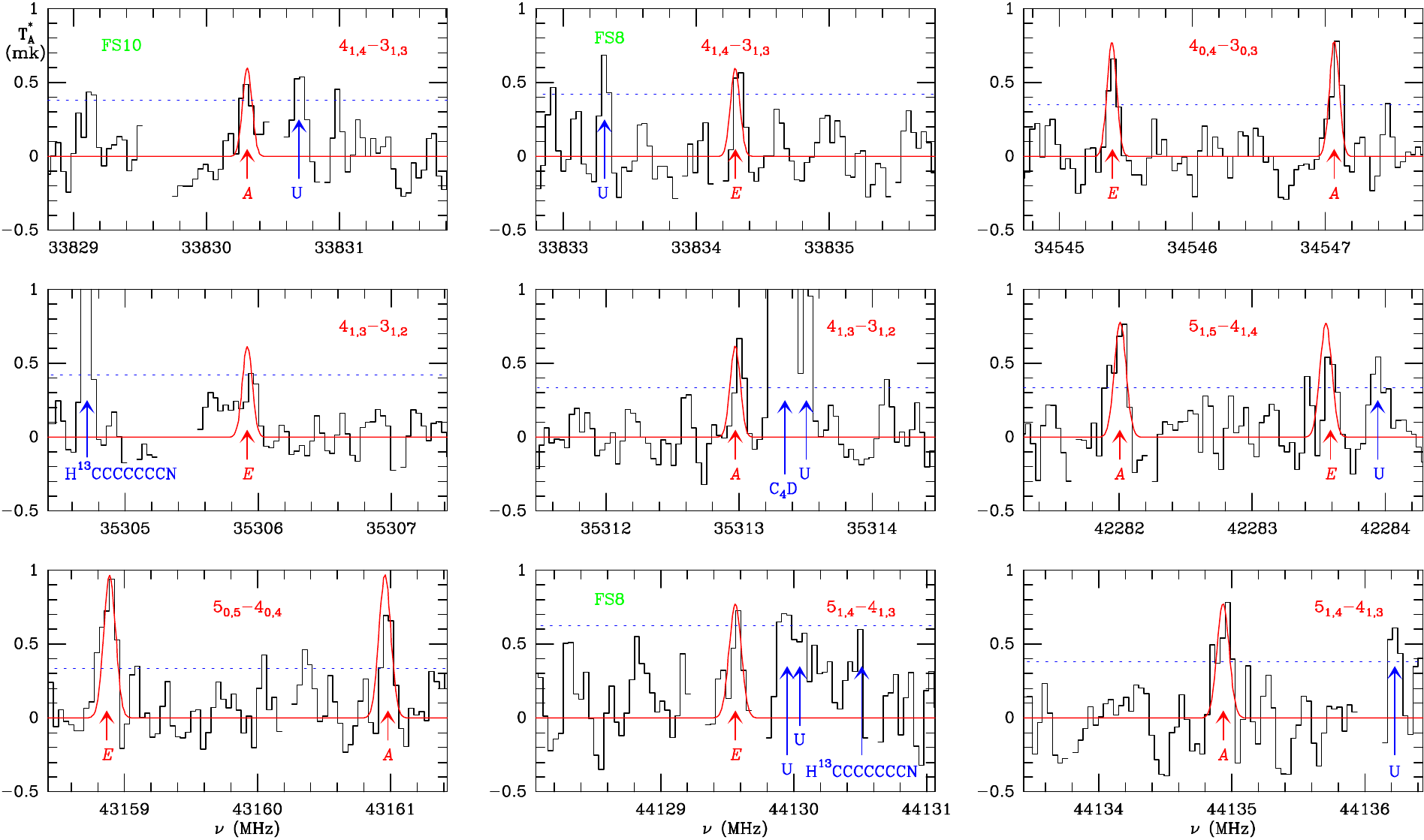}
\caption{Observed transitions with $K_a$ = 0,1 of CH$_3$CHCO in TMC-1.
The abscissa corresponds to the rest frequency of the lines. Frequencies and intensities for the observed lines
are given in Table \ref{obs_line_parameters}.
The ordinate is the antenna temperature, corrected for atmospheric and telescope losses, in millikelvin.
The quantum numbers of each transition are indicated
in the corresponding panel. The red line shows the computed synthetic spectrum for this species for T$_{rot}$=9 K and
a column density of 1.5$\times$\once. Blanked channels correspond to negative features 
produced in the folding of the frequency-switching data. Green labels indicate the transitions for which only one of the frequency-switching data points has been used (FS10
and FS8 correspond to a throw of 10 and 8 MHz, respectively). The dotted line in each panel indicates the 3$\sigma$ value.
}
\label{k0-1}
\end{figure*}

\section{Introduction}

The QUIJOTE\footnote{\textbf{Q}-band \textbf{U}ltrasensitive \textbf{I}nspection \textbf{J}ourney to the \textbf{O}bscure \textbf{T}MC-1 \textbf{E}nvironment} line survey \citep{Cernicharo2021a} toward TMC-1 performed in recent years with the Yebes 40m radio telescope has allowed us to detect more than 40 new molecules in space. This  underlines the importance of this source for a deep understanding of the different chemical processes in cold dense cores.

Among the latest discoveries, there are several cycles such as cyclopentadiene, indene, ortho-benzyne, or fulvenallene  \citep{Cernicharo2021a,Cernicharo2021b,Cernicharo2022}. Cyano and ethynyl derivatives of cyclopentadiene \citep{McCarthy2021,Lee2021,Cernicharo2021c} and cyano derivatives of benzene, naphthalene, and indene \citep{McGuire2018,McGuire2021,Sita2022} have also been detected. We have also detected propargyl \citep{Agundez2021a}, one of the most abundant radicals found. Furthermore, long carbon chains such as vinyl acetylene \citep{Cernicharo2021d}, allenyl acetylene \citep{Cernicharo2021e}, butadiynylallene \citep{Fuentetaja2022}, and ethynylbutatrienylidene \citep{Fuentetaja2022b} have also been discovered toward TMC-1. 
Many of these species were not expected because they did not show a high abundance in chemical models. This   highlights the importance of further study of the dark cloud TMC-1 in order to understand the chemical processes at work in this kind of environment.

Oxygen-bearing complex organic molecules (COMs) are also an important molecular family present in diverse interstellar environments. The star-forming regions, such as Sgr\,B2 and Orion\,KL, are the sources with the highest abundance of COMs.  On the contrary, dark clouds like TMC-1 are characterized  by carbon-rich chemistry, resulting in long carbon chains with low oxygen content. \citet{Agundez2021b} reported the detection of O-bearing species, such as CH$_2$CHCHO, CH$_2$CHOH, HCOOCH$_3$, and CH$_3$OCH$_3$ in TMC-1. Long carbon chain O-bearing molecules with formulae HC$_n$O and C$_n$O (e.g.  HC$_3$O, HC$_7$O, HC$_5$O, and C$_5$O) have also been detected \citep{McGuire2017,Cordiner,Cernicharo2021f}. O-bearing cations, such as HC$_3$O$^+$ and CH$_3$CO$^+$,  have also been recently detected in this source \citep{Cernicharo2020,Cernicharo2021g}.

In the family of C$_3$H$_4$O isomers, the most stable  is $trans$-propenal ($trans$-CH$_2$CHCHO), whose detection was reported by \citet{Agundez2021b}. To date, this is the only isomer in this family detected towards TMC-1. Close in energy, with a difference of 2.8 kJ mol$^{-1}$ compared to $trans$-propenal, is methyl ketene (CH$_3$CHCO). \citet{Bermudez2018} studied this species in different environments and made a theoretical study of the stability of all C$_3$H$_4$O isomers using the coupled cluster (CCSD(T)) ab initio method and the aug-cc-pVTZ basis set. The next member in the series is $cis$-propenal. It is also close in energy to CH$_3$CHCO, with a difference of 5.8 kJ mol$^{-1}$, and therefore it is a candidate for future detection in TMC-1. The last isomer we refer to is cyclopropanone ($c$-H$_4$C$_3$O). Its energy is considerably higher than that of the previously named molecules, specifically 78.1 kJ mol$^{-1}$ with respect to $cis$-propenal (1 kJ mol$^{-1}$=120.3 K), but it is a candidate for detection because the similar and smaller species cyclopropenone ($c$-H$_2$C$_3$O) has been reported by \citet{Loison2016}.

In this letter we report the first clear detection of CH$_3$CHCO (methyl ketene) in TMC-1 using the line survey QUIJOTE \citep{Cernicharo2021a} performed with the Yebes 40m telescope. The formation of this species is investigated in detail using state-of-the-art gas-phase chemical models.

\section{Observations} \label{observations}

The observational data used in this work are part of QUIJOTE, a spectral line survey of TMC-1 in the Q band carried out with the Yebes 40m telescope at the position $\alpha_{J2000}=4^{\rm h} 41^{\rm  m} 41.9^{\rm s}$ and $\delta_{J2000}=
+25^\circ 41' 27.0''$. The receiver was built within the Nanocosmos project\footnote{\texttt{https://nanocosmos.iff.csic.es/}} and consists of two cold high-electron mobility transistor amplifiers covering the 31.0--50.3 GHz band with horizontal and vertical polarizations. Receiver temperatures achieved in the 2019 and 2020 runs vary from 22 K at 32 GHz to 42 K at 50 GHz. Some power adaptation in the down-conversion chains have reduced the receiver temperatures during 2021 to 16\,K at 32 GHz and 30\,K at 50 GHz. The backends are $2\times8\times2.5$ GHz fast Fourier transform spectrometers with a spectral resolution of 38.15 kHz, providing the whole coverage of the Q band in both polarizations.  A more detailed description of the system is given by \citet{Tercero2021}.

The QUIJOTE line survey was carried out in several observing runs between December 2019 and May  2022.  All observations are performed using frequency-switching observing mode with a frequency throw of 8 and 10 MHz. The total observing time on the source for data taken with frequency throws of 8 MHz and 10 MHz is 293 and 253 hours, respectively. Hence, the total observing time on source by May 2022 is 546 hours. The measured sensitivity varies between 0.12 mK at 32 GHz and 0.25 mK at 49.5 GHz.  The sensitivity of QUIJOTE is around 50 times better than that of previous line surveys in the  Q band of TMC-1 \citep{Kaifu2004}. For each frequency throw, different local oscillator frequencies were used in order to remove possible side band effects in the down conversion chain. A detailed description of the QUIJOTE line survey is provided in \citet{Cernicharo2021a}.

The main beam efficiency varies from 0.6 at 32 GHz to 0.43 at 50 GHz \citep{Tercero2021}. The telescope beam size is 56$''$ and 31$''$ at 31 and 50 GHz, respectively.
The intensity scale used in this work, antenna temperature ($T_A^*$), was calibrated using two absorbers at different temperatures and the atmospheric transmission model (ATM) \citep{Cernicharo1985, Pardo2001}. The calibration uncertainties   adopted were  10~\%. All data were analysed using the GILDAS package\footnote{\texttt{http://www.iram.fr/IRAMFR/GILDAS}}.

\begin{figure*}
\centering
\includegraphics[width=0.8\textwidth]{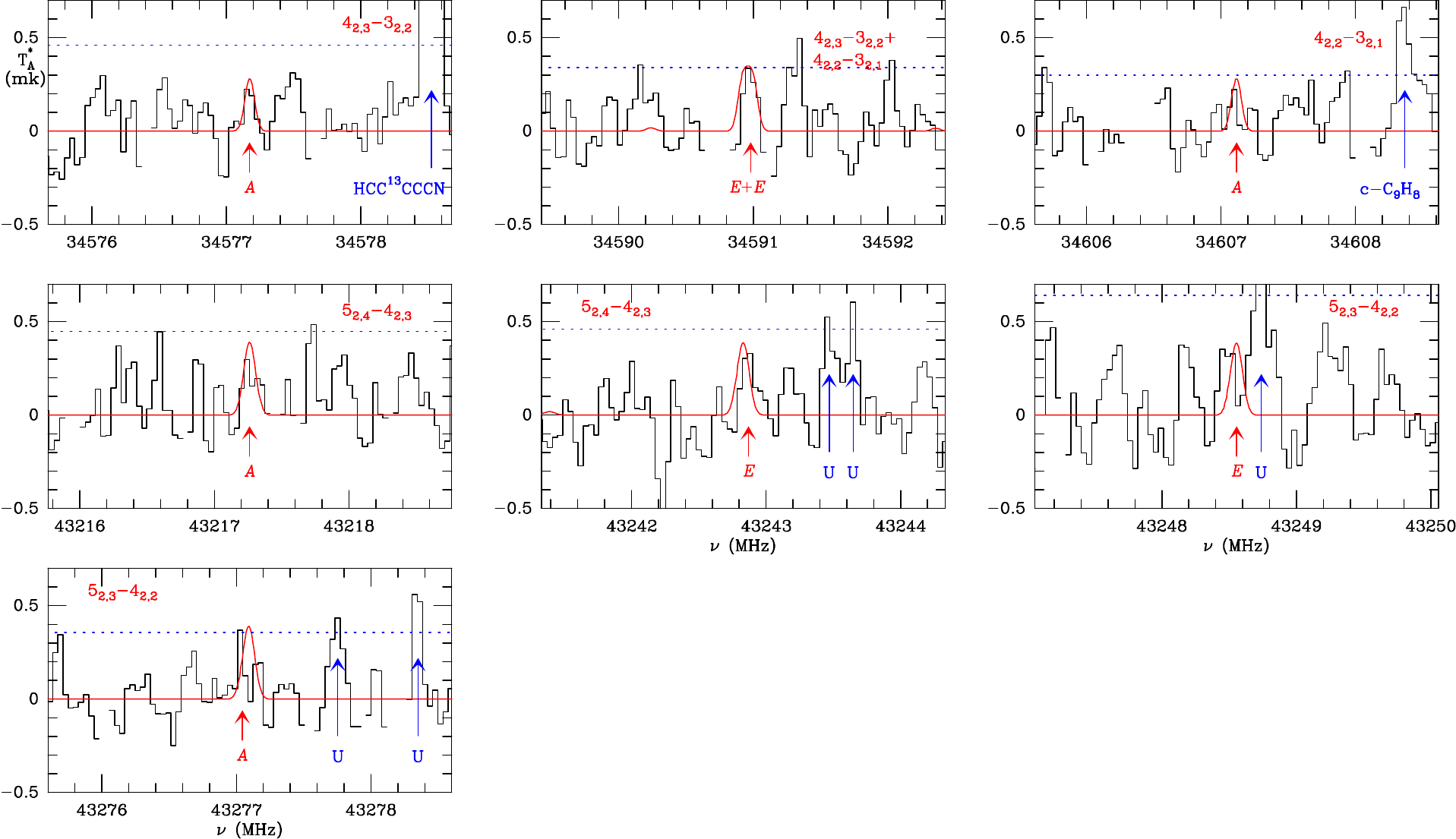}
\caption{Observed transitions with $K_a$ = 2 of CH$_3$CHCO in TMC-1.
The abscissa corresponds to the rest frequency of the lines. Frequencies and intensities for all observed lines
are given in Table \ref{obs_line_parameters}.
The ordinate is the antenna temperature, corrected for atmospheric and telescope losses, in milli Kelvin.
The quantum numbers of each transition are indicated
in the corresponding panel. The red line shows the computed synthetic spectrum for this species for T$_{rot}$=9 K and
a column density of 1.5$\times$\once. Blanked channels correspond to negative features produced in the folding of the frequency-switching data. The dotted line in each panel indicates the 3$\sigma$ value.
}
\label{k2}
\end{figure*}

\section{Results}
\label{results}
Methyl ketene is a nearly prolate molecule having a planar molecular skeleton ($C_s$ frame) with only the two hydrogen atoms of the methyl group out of the plane ($C_{3v}$ top), 
and thus, due to the internal rotation of the methyl top, it belongs to the $G_6$  symmetry group.
As  was shown in the previous rotational analysis of methyl ketene \citep{Bermudez2018}, its $\rho$ parameter, which accounts for the coupling of the methyl torsion with the overall rotation of the molecule, is relatively high 
($\rho = 0.196$). For this reason, it was necessary to employ a Hamiltonian aligned to the $\rho$ vector ($\rho$-axis-method, RAM) that accounts for the coupling of
the two movements. This employed method is incorporated in the RAM36 software \citep[Rho-axis method for 3- and 6-fold barriers;][]{RAM36}. 
The previous model for CH$_3$CHCO reported by \cite{Bermudez2018} was slightly adapted to account for the laboratory observed transitions from \cite{Bak1966} and \cite{Bermudez2018}. The model presented in this work also contains   the observed transitions in TMC-1 between 32.0 and 50.4 GHz since no lines at those frequencies were accounted for in the previous model. A comparison of the parameters obtained in the previous model \citep{Bermudez2018} with the current parameters is presented in Table \ref{params_fit}.
The parameters of the model have barely changed, showing that the transitions were perfectly incorporated in the fit. Furthermore, for all the observed lines in TMC-1 with 
$K_a = 0$ or $1$,  the predictions lie within the experimental error. Some transitions of higher $K_a$ show slightly higher uncertainty than expected; however, this can be explained by   the difficulties of the model to account for the lower energy transitions, already observed in the previous work. 
These issues are related to the strong coupling of the internal motion and the overall rotation of the molecule, and hence to the high complexity of the model.
 The dipole moments used in this work, $\mu_a$ = 1.65 D and $\mu_b$ = 0.33 D,  were reported by \citet{Bermudez2018}.

The line identification was achieved using the MADEX catalogue \citep{Cernicharo2012}. We detected a total of 11 lines (divided into $A$ and $E$ components, due to an internal rotation of the methyl group) within the Q band, together with eight lines having an intensity lower than 3$\sigma$. The intensities range from 0.26 to 0.9 mK. The quantum numbers involved range from $J$ = 3 to $J$ = 5 and $K_a$~$\leq$ 2. 
The derived line parameters are given in Table \ref{obs_line_parameters}. To obtain the column density, we  used a model line fitting procedure, with the  LTE approach for the thin optical lines (see  e.g. \citealt{Cernicharo2021d}). We obtained $N$(CH$_3$CHCO)=1.5$\times$10$^{11}$ cm$^{-2}$ with a rotational temperature of 9 K. The models predict the line intensities in antenna temperature
taking into account the assumed size of the source to correct
for beam dilution, and the beam efficiency of the telescope
at the different frequencies of the observations. We  assume a source of uniform brightness with
a diameter of 80$''$ \citep{Fosse2001}. The H$_2$ column density for TMC-1 is 10$^{22}$
cm$^{-2}$ \citep{Cernicharo1987}, so the abundance of CH$_3$CHCO is 1.5$\times$10$^{-11}$. The predicted synthetic lines for these data are shown in Fig. \ref{k0-1} for $K_a$ = 0,1 and in Fig. \ref{k2} for $K_a$ = 2.

There are several molecules  related to CH$_3$CHCO, so it is interesting to compare their abundances. The most obvious is $trans$-propenal, which is  its more stable isomer. The abundance ratio CH$_2$CHCHO/CH$_3$CHCO is 1.5. This means that  the abundance ratio between the two most stable isomers of the C$_3$H$_4$O family is similar to that of the two most stable isomers of the C$_2$H$_4$O family, in which case C$_2$H$_3$OH/CH$_3$CHO $\sim$1 \citep{Agundez2021b}. 
We can also compare CH$_3$CHCO with ketene, one of the most abundant O-bearing molecules in TMC-1, with an
abundance of 1.4x10$^{-9}$ relative to H$_2$, reported by \citet{Cernicharo2020}. This gives an abundance ratio of CH$_2$CO/CH$_3$CHCO$\sim$93, which means   that  the methylated form of ketene is about two orders of magnitude less abundant than ketene itself. Finally, we compared the abundance of methyl ketene with acetaldehyde (CH$_3$CHO), which has an abundance of 3.5x10$^{-10}$ reported by \citet{Cernicharo2020}. This gives an abundance ratio of CH$_3$CHO/CH$_3$CHCO$\sim$23. 

\begin{small}
        \begin{table*}  
                \caption{Ground torsional state molecular parameters of methyl ketene obtained from the fit using the  RAM36 program ($A$-reduction, $I^{r}$-representation).
                \label{params_fit}}
        \centering       
        \small
        \begin{tabular}{{llcrr}}                                                
                \hline
                Parameter\tablefootmark{a}             & Operator\tablefootmark{b} & $n_{op}t_{op}r_{op}$\tablefootmark{c} & Previous work (MHz)\tablefootmark{d}\tablefootmark{j} &  This work (MHz)\tablefootmark{d}\tablefootmark{k}  \\
                \hline
                $A_{RAM}         $     & $  J^{2}_{a}                                       $  & $2_{0,2}$  &  $       38587.00(11)                   $ & $   38586.91(11)             $    \\         
                $B_{RAM}         $     & $  J^{2}_{b}                                       $  & $2_{0,2}$  &  $        4773.813(29)                  $ & $    4773.833(27)            $    \\        
                $C_{RAM}         $     & $  J^{2}_{c}                                       $  & $2_{0,2}$  &  $        4139.199(46)                  $ & $    4139.249(39)            $    \\        
                $D_{ab}          $     & $  \{J_{a},J_{b}\}                                 $  & $2_{0,2}$  &  $        -3015.26(39)                  $ & $   -3015.62(35)             $    \\        
                $\rho\tablefootmark{e}$& $  p_{\alpha}J_{a}                                 $  & $2_{1,1}$  &  $        0.194354(13)\tablefootmark{e} $ & $   0.194364(12)\tablefootmark{e}$\\        
                $F\tablefootmark{f}$   & $  p^{2}_{\alpha}                                  $  & $2_{2,0}$  &  $          6.4215(13)\tablefootmark{f} $ & $   6.419877(84)\tablefootmark{f}$\\        
                $V_{3}\tablefootmark{f}$& $  \frac{1}{2}(1-\cos3\alpha)                    $  & $2_{2,0}$  &  $         429.38(11)\tablefootmark{f}  $ & $ 429.2326(32)\tablefootmark{f}  $\\        
                $\Delta_{J}      $     & $  -J^{4}                                          $  & $4_{0,4}$  &  $    0.46682(81)~10^{-2}               $ & $   0.46730(77)~10^{-2}      $    \\        
                $\Delta_{JK}     $     & $  -J^{2}J^{2}_{a}                                 $  & $4_{0,4}$  &  $   -0.17168(62)~10^{ 0}               $ & $  -0.17161(59)~10^{ 0}      $    \\        
                $\Delta_{K}      $     & $  -J^{4}_{a}                                      $  & $4_{0,4}$  &  $    0.47291(60)~10^{ 1}               $ & $   0.47296(59)~10^{ 1}      $    \\        
                $\delta_{J}      $     & $  -2J^{2}(J^{2}_{b}-J^{2}_{c})                    $  & $4_{0,4}$  &  $    0.15904(40)~10^{-2}               $ & $   0.15927(38)~10^{-2}      $    \\        
                $\delta_{K}      $     & $  -2\{J^{2}_{a},(J^{2}_{b}-J^{2}_{c})\}           $  & $4_{0,4}$  &  $     0.2120(19)~10^{-1}               $ & $    0.2120(19)~10^{-1}      $    \\        
                $D_{abJ}         $     & $  \{J_{a},J_{b}\}J^{2}                            $  & $4_{0,4}$  &  $     0.3983(19)~10^{-1}               $ & $    0.3993(18)~10^{-1}      $    \\        
                $D_{abK}         $     & $  \{J_{a},J_{b}\}J^{2}_{a}                        $  & $4_{0,4}$  &  $     0.5329(38)~10^{ 0}               $ & $    0.5320(36)~10^{ 0}      $    \\        
                $\rho_{J}        $     & $  p_{\alpha}J_{a}J^{2}                            $  & $4_{1,3}$  &  $    -0.4253(14)~10^{ 0}               $ & $   -0.4254(13)~10^{ 0}      $    \\        
                $\rho_{K}        $     & $  p_{\alpha}J^{3}_{a}                             $  & $4_{1,3}$  &  $    0.21648(59)~10^{ 2}               $ & $   0.21687(53)~10^{ 2}      $    \\        
                $F_{J}           $     & $  p^{2}_{\alpha}J^{2}                             $  & $4_{2,2}$  &  $    0.15159(31)~10^{ 0}               $ & $   0.15136(26)~10^{ 0}      $    \\        
                $F_{K}           $     & $  p^{2}_{\alpha}J^{2}_{a}                         $  & $4_{2,2}$  &  $    -0.4535(31)~10^{ 2}               $ & $   -0.4580(14)~10^{ 2}      $    \\        
                $F_{ab}          $     & $  \frac{1}{2}\{J_{a},J_{b}\}p^{2}_{\alpha}       $  & $4_{2,2}$  &  $      0.902(35)~10^{ 0}               $ & $     0.898(30)~10^{ 0}      $    \\        
                $V_{3J}          $     & $  J^{2}(1-\cos3\alpha)                            $  & $4_{2,2}$  &  $   -0.13759(47)~10^{ 2}               $ & $  -0.13807(40)~10^{ 2}      $    \\        
                $V_{3K}          $     & $  J^{2}_{a}(1-\cos3\alpha)                        $  & $4_{2,2}$  &  $    -0.1229(28)~10^{ 3}               $ & $   -0.1246(27)~10^{ 3}      $    \\        
                $V_{3ab}         $     & $  \frac{1}{2}(\{J_{a},J_{b}\})(1-\cos3\alpha)    $  & $4_{2,2}$  &  $     -0.647(15)~10^{ 2}               $ & $    -0.647(14)~10^{ 2}      $    \\        
                $V_{3bc}         $     & $  (J^{2}_{b}-J^{2}_{c})(1-\cos3\alpha)            $  & $4_{2,2}$  &  $     0.2134(47)~10^{ 1}               $ & $    0.2184(39)~10^{ 1}      $    \\        
                $D_{3ac}         $     & $  \frac{1}{2}\{J_{a},J_{c}\}\sin3\alpha          $  & $4_{2,2}$  &  $     0.4186(50)~10^{ 3}               $ & $    0.4193(49)~10^{ 3}      $    \\        
                $\rho_{m}        $     & $  p^{3}_{\alpha}J_{a}                             $  & $4_{1,3}$  &  $      0.514(11)~10^{ 2}               $ & $    0.5324(15)~10^{ 2}      $    \\        
                $F_{m}           $     & $  p^{4}_{\alpha}                                  $  & $4_{4,0}$  &  $     -0.443(15)~10^{ 2}               $ & $  -0.46728(72)~10^{ 2}      $    \\        
                $V_{6}           $     & $  \frac{1}{2}(1-\cos6\alpha)                     $  & $4_{4,0}$  &  $    -0.3088(44)~10^{ 6}               $ & $  -0.30159(12)~10^{ 6}      $    \\        
                $\Phi_{J}        $     & $  J^{6}                                           $  & $6_{0,6}$  &  $    0.21428(94)~10^{-7}               $ & $   0.21471(90)~10^{-7}      $    \\        
                $\Phi_{JKK}      $     & $  J^{2}J^{4}_{a}                                  $  & $6_{0,6}$  &  $    -0.7678(62)~10^{-4}               $ & $   -0.7686(60)~10^{-4}      $    \\        
                $\Phi_{K}        $     & $  J^{6}_{a}                                       $  & $6_{0,6}$  &  $     0.8161(75)~10^{-3}               $ & $    0.8154(73)~10^{-3}      $    \\        
                $\phi_{J}        $     & $  2J^{4}(J^{2}_{b}-J^{2}_{c})                     $  & $6_{0,6}$  &  $     0.9235(85)~10^{-8}               $ & $   0.9245(80)~10^{-8}$    \\        
                $\rho_{JK}       $     & $  p_{\alpha}J^{3}_{a}J^{2}                        $  & $6_{1,5}$  &  $     0.5936(53)~10^{-3}               $ & $    0.5941(51)~10^{-3}      $    \\        
                $\rho_{KK}       $     & $  p_{\alpha}J^{5}_{a}                             $  & $6_{1,5}$  &  $    -0.6917(54)~10^{-2}               $ & $   -0.6895(52)~10^{-2}      $    \\        
                $F_{JK}          $     & $  p^{2}_{\alpha}J^{2}J^{2}_{a}                    $  & $6_{2,4}$  &  $    -0.1522(14)~10^{-2}               $ & $   -0.1524(13)~10^{-2}      $    \\        
                $F_{KK}          $     & $  p^{2}_{\alpha}J^{4}_{a}                         $  & $6_{2,4}$  &  $     0.1585(15)~10^{-1}               $ & $    0.1578(15)~10^{-1}      $    \\        
                $V_{3JK}         $     & $  J^{2}J^{2}_{a}(1-\cos3\alpha)                   $  & $6_{2,4}$  &  $     0.1922(51)~10^{-1}               $ & $    0.1954(46)~10^{-1}      $    \\        
                $\rho_{mbc}      $     & $  p^{3}_{\alpha}J_{a}(J^{2}_{b}-J^{2}_{c})        $  & $6_{3,3}$  &  $      0.677(15)~10^{-3}               $ & $     0.671(14)~10^{-3}      $    \\        
                $V_{6ab}         $     & $  \{J_{a},J_{b}\}(1-\cos6\alpha)                  $  & $6_{4,2}$  &  $      0.425(10)~10^{ 2}               $ & $    0.4207(97)~10^{ 2}      $    \\  \hline                                                                                                                                                                                                 
                $\sigma_{\text{rms}/\text{wrms}}\tablefootmark{g}$ &                           &            &  $     4.42~10^{-2}/0.74~~              $ & $    4.18~10^{-2}/0.73~~     $    \\        
                $Jmax/K_{a}max\tablefootmark{h}$ &                                             &            &  $      38/18~~~~~~~                    $ & $     38/18~~~~~~~           $    \\        
                $N_{lines}\tablefootmark{i}$ &                                                 &            &  $      3040~~~~~~~~                    $ & $      3059~~~~~~~~           $    \\ \hline 
\end{tabular}                                               
\tablefoot{
        \tablefoottext{a}{Parameter nomenclature.}
        \tablefoottext{b}{Torsional-rotation operators employed in the model in the $I^{r}$ representation.}
        \tablefoottext{c}{ $n_{op}$ is the total order operator, which is $n_{op}=t_{op}+r_{op}$, with $t_{op}$ the order of the torsional part and $r_{op}$ that of the rotational part.}
        \tablefoottext{d}{Unless indicated, all constants are expressed in MHz.}
        \tablefoottext{e}{$\rho$ value is unitless.}
        \tablefoottext{f}{$F$ and $V_{3}$ values are expressed in $cm^{-1}$.}
        \tablefoottext{g}{Fit root mean square error in MHz / weighted root mean square. }
        \tablefoottext{h}{Maximum value of quantum number $J$ and $K_{a}$ included in the fit.}   
        \tablefoottext{i}{Number of transitions included in the fit.}
        \tablefoottext{j}{Ground torsional state molecular parameters from \citet{Bermudez2018}.}
        \tablefoottext{k}{Ground torsional state molecular parameters derived from  this work.}
}
\end{table*}    

\end{small}     
\normalsize

\section{Chemical model} 
\label{discussion}

\begin{figure} \label{fig:abun}
\centering
\includegraphics[width=0.9\columnwidth]{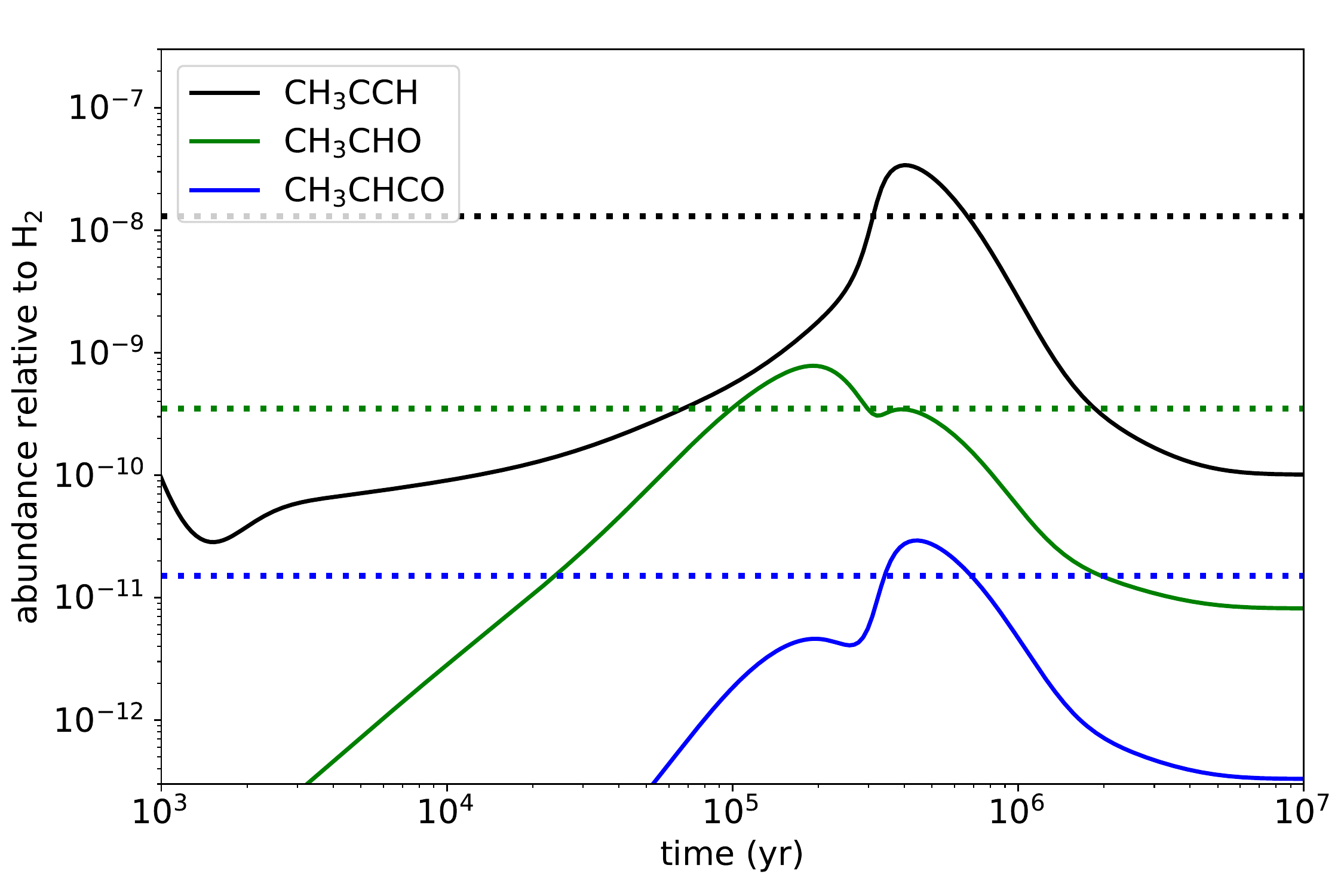}
\caption{Calculated fractional abundance of CH$_3$CHCO and its relevant precursors CH$_3$CCH and CH$_3$CHO as a function of time. The horizontal dotted line corresponds to the observed abundance in TMC-1.}
\end{figure}

To investigate the formation of methyl ketene in TMC-1 we  carried out gas-phase chemical modelling calculations. The model parameters and chemical network are the same used in \cite{Cernicharo2021g} to model the chemistry of O-bearing molecules following the discovery of HC$_3$O and C$_5$O. We   added CH$_3$CHCO as a new species, with a simple chemical scheme of formation and destruction. Although some chemical models  (e.g. \citealt{Garrod2022}) indicate that grain surface processes can help to explain the presence of some complex organic molecules (e.g. methyl formate and dimethyl ether;  \citealt{Agundez2021b}) in cold sources such as TMC-1, here we aim to evaluate whether purely gas phase processes can account for the formation of CH$_3$CHCO in TMC-1.

Methyl ketene is not included in the UMIST\footnote{\texttt{http://udfa.ajmarkwick.net/}} \citep{McElroy2013} or KIDA\footnote{\texttt{https://kida.astrochem-tools.org/}} \citep{Wakelam2015} databases, although some information is available in the NIST Chemical Kinetics database\footnote{\texttt{https://kinetics.nist.gov/}}. There are several plausible reactions of formation. The reaction between OH and CH$_3$CCH is a potential source of methyl ketene because it is relatively rapid at low temperatures, with a measured rate coefficient of 5.08\,$\times$\,10$^{-12}$ cm$^3$ s$^{-1}$ at 69 K \citep{Taylor2008}. However, no information is available on the product distribution, and thus here we assume that CH$_3$CHCO and  H are the main products. The reaction between OH and the non-polar isomer allene (CH$_2$CCH$_2$) has also been   found to be rapid at low temperatures, although the main products seem to be H$_2$CCO  and  CH$_3$ \citep{Daranlot2012}, and we thus do not include it as a source of methyl ketene. Another reaction that can provide an efficient formation route to methyl ketene is CH + CH$_3$CHO. This reaction was   studied by \cite{Goulay2012}, who found that CH$_3$CHCO is formed with a branching ratio of 0.39. However, the rate coefficient  has not been measured, although \cite{Wang2017} studied the reaction theoretically and found that the formation of CH$_3$CHCO is barrierless. We thus adopted a rate coefficient of 2.41\,$\times$\,10$^{-10}$ cm$^3$ s$^{-1}$, as measured for CH and  H$_2$CO \citep{Hancock1992}, and the branching ratio measured by \cite{Goulay2012}.

There are other reactions that could potentially form CH$_3$CHCO, although they are unlikely to be efficient in TMC-1. For example, the reaction between HCO and C$_2$H$_4$ has a barrier \citep{Lesclaux1986,Xie2005a}, and the same happens for the reaction between C$_2$H$_3$ and H$_2$CO \citep{Xie2005b}. The reaction between CH$_3$ and  H$_2$CCO could provide a simple pathway to CH$_3$CHCO by simply substituting one H atom of ketene by a methyl group, although ketene does not show a high reactivity with radicals. \cite{Semenikhin2018} studied theoretically the reaction between CH$_3$ and H$_2$CCO;  although they did not consider the formation of CH$_3$CHCO and H, they found that all the explored channels have barriers. Therefore, we did not include this reaction. Finally, we considered that in TMC-1 methyl ketene is mostly destroyed through reactions with the most abundant cations, such as C$^+$, HCO$^+$, and H$_3$O$^+$.

The fractional abundance calculated for CH$_3$CHCO is shown in Fig.~\ref{fig:abun} as a function of time. It is seen that the peak abundance, reached at a time of some 10$^5$ yr, agrees very well with the abundance derived from the observations. The two formation reactions considered here (i.e. OH + CH$_3$CCH and CH + CH$_3$CHO) contribute to the formation of methyl ketene. The abundances calculated for CH$_3$CCH and CH$_3$CHO are in good agreement with those obtained in \citet{Cabezas2021} and \citet{Cernicharo2020}. Further research on the low temperature kinetics and the product distribution of these two reactions will be of great interest to shed light on the origin of methyl ketene in TMC-1.

\section{Conclusions}
We have presented the first detection of CH$_3$CHCO towards TMC-1. We used the QUIJOTE line survey taken with the Yebes 40m radiotelescope, with which we observed a total of 11 lines with an intensity higher than 3$\sigma$ and another 8 lines with an intensity lower than 3$\sigma$, involving $J$ = 3 to $J$ = 5 and $K_a$ $\leq$ 2. The rotational temperature is 9 K and the derived column density $N$(CH$_3$CHCO)=1.5$\times$10$^{11}$ cm$^{-2}$. These results imply that methyl ketene is 1.46 less abundant
than its most stable isomer (trans-CH$_2$CHCHO), a value quite similar to the abundance ratio of the isomers vinyl alcohol to acetaldehyde of 1. The observed abundance of methyl ketene is well explained using our gas phase chemical model, considering the formation reactions from propyne and acetaldehyde, and the destruction reactions with the most abundant cations of TMC-1. 

\begin{acknowledgements}

We thank Ministerio de Ciencia e Innovaci\'on of Spain (MICIU) for funding support through projects
PID2019-106110GB-I00, PID2019-107115GB-C21 / AEI / 10.13039/501100011033, and
PID2019-106235GB-I00. We also thank ERC for funding
through grant ERC-2013-Syg-610256-NANOCOSMOS. C.B. thanks Ministerio de Universidades for her "Maria Zambrano" grant at UVa (CONVREC-2021-317).
\end{acknowledgements}

\normalsize

\onecolumn
\begin{appendix}

\section{Observed line parameters}
\label{sec:rotaspctro} 

The line parameters derived for this work were obtained by fitting a Gaussian line profile to the observed data, using the software Class (GILDAS package). We use a window of $\pm$ 15 km s$^{-1}$ around the  V$_{LSR}$ (5.83km s$^{-1}$) of the source for each transition. The results are given in Table \ref{obs_line_parameters}. The observed lines of methyl ketene are shown in Fig. \ref{k0-1} for $K_a$ = 0,1 and in Fig. \ref{k2} for $K_a$ = 2.

\begin{table}[h]
\caption{Observed line parameters for CH$_3$CHCO} \label{obs_line_parameters}
\centering
\begin{tabular}{ccccccc}
\hline 
\hline
Transition$^1$     &$\nu_{obs}$~$^a$ & $\nu_{obs}$-$\nu_{cal}$   & $\int T_A^* dv$~$^b$&  $\Delta$v$^c$    & $T_A^*$~$^d$ & Notes  \\
                   &  (MHz)    & (MHz)        & (mK\,km\,s$^{-1}$)  &  \multicolumn{1}{c}{(km\,s$^{-1}$)}    & \multicolumn{1}{c}{(mK)}        \\

\hline

4$_{1,4}$-3$_{1,3}$~$A$ &  33830.302$\pm$0.010 &  -0.013   & 0.51$\pm$0.13 &   0.98$\pm$ 0.31 &     0.49$\pm$0.13 & A \\ 
4$_{1,4}$-3$_{1,3}$~$E$  &  33834.322$\pm$0.010 & 0.027    & 0.49$\pm$0.12 &   0.66$\pm$ 0.18 &    0.69$\pm$0.14 & B \\ 
4$_{0,4}$-3$_{0,3}$~$E$ &  34545.403$\pm$0.010 &  0.004   & 0.58$\pm$0.10 &   0.75$\pm$ 0.14 &     0.72$\pm$0.12 \\
4$_{0,4}$-3$_{0,3}$~$A$  &  34547.083$\pm$0.010 & 0.007    & 0.69$\pm$0.11 &   0.82$\pm$ 0.17 &     0.78$\pm$0.12 \\
4$_{1,3}$-3$_{1,2}$~$E$ &  35305.954$\pm$0.022 &  0.029   & 0.38$\pm$0.15 &   0.76$\pm$ 0.35 &     0.46$\pm$0.14 \\
4$_{1,3}$-3$_{1,2}$~$A$  &  35313.007$\pm$0.010 & 0.026    & 0.51$\pm$0.10 &   0.69$\pm$ 0.15 &     0.70$\pm$0.11 \\
5$_{1,5}$-4$_{1,4}$~$A$ &  42282.025$\pm$0.011 &   0.007  & 0.46$\pm$0.16 &   0.50$\pm$ 0.18 &     0.86$\pm$0.11 \\
5$_{1,5}$-4$_{1,4}$~$E$  &  42283.591$\pm$0.013 & 0.032    & 0.40$\pm$0.11 &   0.64$\pm$ 0.19 &     0.59$\pm$0.11 \\
5$_{0,5}$-4$_{0,4}$~$E$ &  43158.875$\pm$0.011 &  -0.020   & 0.88$\pm$0.14 &   0.93$\pm$ 0.16 &     0.89$\pm$0.13 \\
5$_{0,5}$-4$_{0,4}$~$A$  &  43160.982$\pm$0.010 & 0.011    & 0.51$\pm$0.11 &   0.60$\pm$ 0.13 &     0.79$\pm$0.13 \\
5$_{1,4}$-4$_{1,3}$~$E$ &  44129.579$\pm$0.012 &  0.07   & 0.33$\pm$0.12 &   0.48$\pm$ 0.19 &     0.65$\pm$0.21 & A \\
5$_{1,4}$-4$_{1,3}$~$A$  &  44134.955$\pm$0.014 & 0.07    & 0.53$\pm$0.14 &   0.63$\pm$ 0.18 &     0.78$\pm$0.13 \\
\hline 
4$_{2,3}$-3$_{2,2}$~$A$ &  34577.168$\pm$0.011 &  -0.012   & 0.10$\pm$0.08 &   0.33$\pm$ 3.42 &     0.28$\pm$0.16 \\
4$_{2,3}$-3$_{2,2}$~$E$ + 4$_{2,2}$-3$_{2,1}$~$E$  &  34590.976$\pm$0.015 & 0.043 \& -0.021    & 0.42$\pm$0.13 &   0.91$\pm$ 0.26 &     0.43$\pm$0.12 & C \\
4$_{2,2}$-3$_{2,1}$~$A$ &  34607.086$\pm$0.010 &  -0.035   & 0.11$\pm$0.07 &   0.35$\pm$ 1.80 &     0.28$\pm$0.10 \\
5$_{2,4}$-4$_{2,3}$~$A$  &  43217.244$\pm$0.028 & -0.026    & 0.13$\pm$0.10 &   0.44$\pm$ 0.34 &     0.26$\pm$0.15 \\
5$_{2,4}$-4$_{2,3}$~$E$ &  43242.864$\pm$0.019 &  0.026   & 0.19$\pm$0.09 &   0.58$\pm$ 0.29 &     0.31$\pm$0.16 \\
5$_{2,3}$-4$_{2,2}$~$E$  &  43248.485$\pm$0.033 & -0.074    & 0.21$\pm$0.13 &   0.65$\pm$ 0.35 &     0.31$\pm$0.22 \\
5$_{2,3}$-4$_{2,2}$~$A$ &  43277.034$\pm$0.025 & -0.071    & 0.12$\pm$0.09 &   0.32$\pm$ 0.39 &     0.36$\pm$0.12 \\
\hline
\end{tabular}

\tablefoot{\\
\tablefoottext{1}{Quantum numbers are $J'_{K'_{a,}K'_{c}}$ - $J_{K_{a,}K_{c}}$.}\\
\tablefoottext{a}{Observed frequency of the transition assuming a LSR velocity of 5.83 km s$^{-1}$.}\\
\tablefoottext{b}{Integrated line intensity in mK\,km\,s$^{-1}$.}\\
\tablefoottext{c}{Linewidth at half intensity derived by fitting a Gaussian function to
the observed line profile (in km\,s$^{-1}$).}\\
\tablefoottext{d}{Antenna temperature in millikelvin.}\\
\tablefoottext{A}{Frequency switching data with a throw of 10 MHz only. Negative feature present in the data with a 8 MHz throw.}\\
\tablefoottext{B}{Frequency switching data with a throw of 8 MHz only. Negative feature present in the data with a 10 MHz throw.}\\
\tablefoottext{C}{The line is blended and corresponds to two transitions}\\
}
\end{table}

\normalsize
\end{appendix}

\end{document}